# Implementing Smart Contracts: The case of NFT-rental with "pay-per-like"

**Research Paper**

Alfred Sopi, Johannes Schneider, Jan vom Brocke

University of Liechtenstein, Vaduz, Liechtenstein
{alfred.sopi, johannes.schneider, jan.vombrocke}@uni.li

**Abstract.** Non-fungible tokens(NFTs) are on the rise. They can represent artworks exhibited for marketing purposes on webpages of companies or online stores - analogously to physical artworks. Lending of NFTs is an attractive form of passive income for owners but comes with risks (e.g., items are not returned) and costs for escrow agents. Similarly, renters have difficulties in anticipating the impact of artworks, e.g., how spectators of NFTs perceive them. To address these challenges, we introduce an NFT rental solution based on a pay-per-like pricing model using blockchain technology, i.e., smart contracts based on the Ethereum chain. We find that blockchain solutions enjoy many advantages also reported for other applications, but interestingly, we also observe dark sides of (large) blockchain fees. Blockchain solutions appear unfair to niche artists and potentially hamper cultural diversity. Furthermore, a trust-cost tradeoff arises to handle fraud caused by manipulation from parties outside the blockchain. All code for the solution is publicly available at: https://github.com/asopi/rental-project

## 1 Introduction

Since the invention of Bitcoin, blockchain technology has rapidly advanced. Adoption of blockchain technology, especially smart contracts, has also been linked with positive performance, e.g., in an inter-organizational setting (Werner *et al.*, 2021). Thus, the time seems more than ripe for widespread adoption. Multiple applications have emerged, and there is a growing body of works that discuss their implementation such as in the energy sector (Kirli *et al.*, 2022), insurance domain(Chondrogiannis *et al.*, 2022), equity market(vom Brocke *et al.*, 2018), and event ticketing(Regner, Urbach and Schweizer, 2019).

However, the full potential of blockchain technology has not yet been explored, in particular, in emerging areas such as digital artworks whose ownership and authenticity can be ensured using non-fungible tokens (NFTs). NFT usage has been growing in the last years (Christodoulou *et al.*, 2022). NFTs might soon be omnipresent.

A non-fungible token is a unique digital identifier that can neither be copied, substituted, nor subdivided (Voshgmir, 2018). It is stored in a blockchain and serves to certify authenticity and ownership. NFTs can be traded or rented. They typically contain references to digital files of various formats, e.g., texts, audio, and images that potentially resemble digital work.

In this work, we focus on an under-explored area, i.e., rental of NFTs with value-based pricing. More specifically, the rental price of artwork is based on the perception of visitors of exhibitions expressed using "likes" as prevalent throughout the digital world, especially, in social media (Ding *et al.*, 2017). NFT renting should enable NFT owners to rent digital assets for temporary use. Especially for art, rental of NFTs is very attractive. For NFT owners, it is lucrative to generate passive income instead of keeping their valuable NFTs locked in their accounts. Renters benefit from the possibility of acquiring the NFT for a limited time, so they do not have to pay high costs compared to purchasing the NFT.

As of now, few organizations are pioneering the NFT rental market, and most solutions have not yet been launched (Beyer, 2022). However, since NFTs are becoming increasingly important and can be very expensive, the need for renting these virtual goods is likely to increase in the upcoming years. Particularly in light of the growing world of virtual assets and the high prices attached to popular NFTs, the need for alternative financing and income-generating methods has increased (J.P.Morgan, 2022).

A rental solution should leverage opportunities of blockchain technology to eliminate typical issues in renting. The most pressing issue for the renter is the return of the item, once the agreed rental period has passed, and being ensured that payments take place. The lender also faces risks. Artwork can trigger a diverse set of reactions, i.e., it can be inspirational, but also trigger disgust or cause no reaction at all. Therefore, it seems advantageous for a renter to pay for positive reactions, e.g., by counting visitors liking the artwork as measured by clicks on a like button. Existing rental solutions are based on a pay-per-use model. They are not able to address fundamental risks of renters or the idea of making payments depending on the reception of an artwork by its audience. To address these issues, we explore novel solutions using state-of-the-art blockchain technology. Our pay-per-like proposal is analogous to the pay-per-click model, which is one of the modern means of online advertising, wherein an advertiser profits based on the number of clicks made by users on a particular website or an app. There are also a number of challenges in implementing such solutions. For example, click-fraud poses a risk. Each click incurs a cost to the advertiser. Rivals might aim at depleting a competitor's budget through fraudulent clicks or websites hosting 3$^{rd}$ party advertisement have an incentive to generate clicks to increase revenues(Zhang and Guan, 2008). Furthermore, a lender might have to pay a collateral, which poses a risk but also implies that assets are bound for the duration of the rental and cannot be used otherwise.

To develop and evaluate our NFT rental solution, we follow a rigorous design science research (DSR) approach based on Peffers *et al.* (2007). We leverage existing knowledge through a literature review as well as practitioners' perspectives through an illustrative scenario as common in IS Research (Vom Brocke *et al.*, 2020). By instantiating our framework in a Proof of Concept (PoC) paired with extensive testing, we demonstrate our approach's feasibility and correctness and evaluate its fitness to solve rental-related problems(Hevner *et al.*, 2004). Especially, we address the requirement that the rental price should be value-based, i.e., determined based on feedback from renters.

The key contributions of the paper are: Conceptualizing, implementing and evaluating a value-based rental solution. We also discuss our (and other) solutions in a general setting with three actors, i.e, lenders, renters and consumers. Especially, we unveil a cost-trust trade-off between the actors and negative aspects of high blockchain fees incurred for decentralized Web3 applications, which might manifest as a disadvantage for less established artists.

The paper is organized as follows: We provide background on technology and pricing models, followed by describing the design science process serving as the methodology for our study. The results covering problem analysis, design objectives and implementation, evaluation are stated prior to the discussion and conclusions.

## 2 Background

**Smart Contracts** are a computerized transaction protocol that executes the terms of a contract (Szabo, 1994). Smart contracts based on a blockchain are transparent, irreversible, and traceable and do not need intermediaries and human decision-making (Antonopoulos and Wood, 2019, pp. 127–131). However, once deployed on the blockchain, smart contracts cannot be changed, emphasizing the importance of careful development. **Ethereum** provides a blockchain that is well-suited for decentralized applications (Dapps) and smart contracts using Ether(ETH) as currency (Buterin, 2014). The Ethereum virtual machine (EVM) executes code in a global blockchain network. It interprets, stores, and executes smart contracts deployed on the blockchain. Transactions can either be contract creation to deploy the smart contract on the blockchain or message calls to interact with the smart contract(Hirai, 2017).

An account on the Ethereum blockchain is an entity with an ETH balance that can execute transactions. **Accounts** can be user-owned or deployed as smart contracts. Both have the ability to receive, hold, and send ETH and interact with deployed smart contracts (Ethereum Community, 2022b). Anyone can generate an unlimited number of accounts, which correspond to a private and public key pair. The blockchain acts as an enormous ledger with transactions being transferred from one account to another.

**Tokens** are abstractions representing the ownership of physical or virtual property. They are cryptographically secured on the blockchain. Ownership is associated with certain privileges, such as the right to use or send that property. The tracking of ownership changes is stored on the blockchain and can be viewed and verified transparently by everyone (Solorio, Kanna and Hoover, 2019, p.23). Tokens are implemented using smart contracts, which typically follow a specific standard called **Ethereum Request for Comments (ERC)**. The ERC-721 introduces the first standard for non-fungible tokens (NFTs) (Ethereum Community, 2022a). The term non-fungible means that the object is unique, not replaceable, and not divisible. An ERC-721 token is used to uniquely identify any content, such as an image, song, or a video. This type of token is used by platforms that offer collectibles, access keys, lottery tickets, etc. Moreover, these unique tokens can be transferred from one account to another. The sender of a transaction must pay a gas fee in ETH depending on computational demands of the transaction (i.e., smart contract) and network utilization(Dr. Wood, 2022,p.8)

A **decentralized application** (Dapp) runs on a peer-to-peer network. Smart contracts are used to implement the backend of such applications, which are then deployed on a blockchain and executed through a frontend or user interface(Zheng *et al.*, 2021,pp.253-280). **Web3** or the decentralized web refers to a distributed internet that aims to eliminate trusted third parties (Antonopoulos and Wood, 2019,p.267). Web2 requires internet users to trust various private companies such as social media platform operators to act in the public's interest. Web3 intends to solve this issue by providing ownership capabilities. The Web3 ecosystem consists of Dapps that adhere to the fundamental principle that users own their data and therefore exert full control over it, unlike applications (and data) managed by centralized organizations.

There are a few **NFT rental** solutions, e.g., IQSpace (2022), Bored Jobs (2022), and reNFT (2022). But none implements an innovative value-based payment solution such as pay-per-click. Typically, the smart contract returns the rented NFT plus the rental fee to the lender after the expiration of the rental period. For collateralized renting, a renter has to deposit a collateral plus the rental fees defined by the lender. The renter then receives the NFT into their account, and the collateral remains in the smart contract, which acts as an escrow account until the renter returns the rented NFT. For collateral-free renting, wrapped NFTs enable the rental of NFTs without a collateral. A wrapped NFT is transferred to the renter's account. Service providers on whose platforms the NFT is to be used can then use the smart contract to verify that the wrapped NFTs originated from a rental deal. For collateral-free, non-wrapped rental the NFT is transferred from the lender's account to the smart contract, the account address from the renter is used to verify and retrieve its rented NFTs.

**Pricing Models:** The Pay-Per-Use Model (PPU) model also known as the pay-as-you-go usage-based pricing (UBP) model, involves sellers offering identical or comparable goods at different prices, depending on the volume of the goods being purchased(Weinhardt *et al.*, 2009). The PPU model associates service units (e.g., units per time) with fix-priced values so that option prices vary with the volume of service units used(Bauer and Wildman, 2012). The customer pays only for the use of the product or service.

The Pay-Per-Click Model (PPC) model is one of the modern means of online advertising, wherein an advertiser profits based on the number of clicks made by users on a particular website or an app. The advertiser pays for each click collected by the ads, rather than allowing the advertiser to pay a fixed price for a spot on a website. Search engines such as Google or Yahoo exemplify sites with a pay-per-click model (Khraim and Alkrableih, 2015, pp.180-181). One of the most significant limitations of the PPC is the risk of click fraud. The literature recognizes click fraud as the intended and ill-conceived use of PPC when advertising to cause financial damage to another party.

Ives (2005) argued that some individuals use click bots or hit bots to generate invalid clicks on an ad to increase the click count. Click fraud may also include unsatisfied users or employees who willingly click on ads to increase the number of clicks, thereby damaging the company's budget. Attackers can also deliberately cause undesired traffic through scams to damage competitors (Rosso and Jansen, 2010).

# 3    Methodology

We employ the DSR process by Peffers et al. (2007). Design science aims to conduct research through the development and evaluation of artifacts designed to meet, identify, and leverage business needs (Hevner et al., 2004, pp. 79–80). Design knowledge generated in DSR, can take the form of design entities, e.g. constructs and methods, or the form of design theories. It can be derived inductively, in that the design focuses on situated solutions to situated problems and is, then, further generalized in order to increase the projectability of the design knowledge(Vom Brocke *et al.*, 2020). Specifically, DSR develops prototypes that are demonstrated through iterative analysis, design, and implementation, whether the planned solution can be effectively and efficiently implemented (see Figure 1). Since DSR supports the generation of knowledge on the design of innovative solutions to real-world problems, DSR has been found particularly suitable in blockchain contexts (Regner, Urbach, and Schweizer, 2019, p.5). Knowledge is collected through various methods in several steps. This enables a systematic observation, evaluation, and refinement of the obtained information.

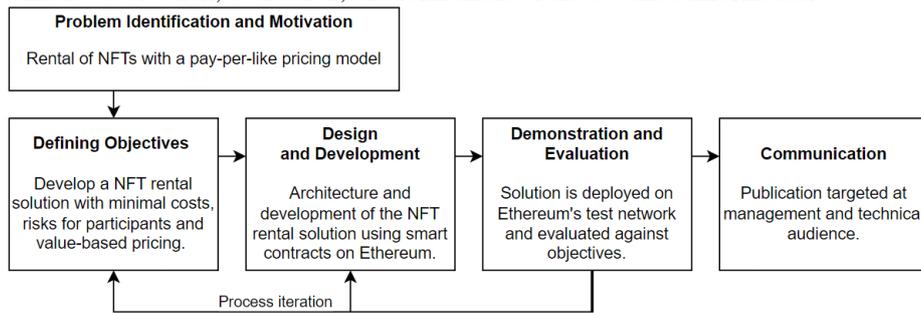

*Figure 1: Design Science Research Methodology for the NFT rental solution*

In the first step, we analyze the problem domain based on a literature review, which includes identification of key concepts and terms, and current implementations of smart contract-based solutions, especially rental based-solutions for NFTs to identify shortcomings and their causes. The findings serve as a basis to derive requirements for a tool to address these challenges. The second step comprises the artifact design, which focuses on the development of the Dapp. In the first evaluation step, the two lead authors of this research interacted with the web-based tool prototype, which led to another development cycle to improve shortcomings. For communication, we target management- and technically-oriented audiences. The former can use schematic UML diagrams and reasoning about benefits and challenges, while for the latter we made source code available at: https://github.com/asopi/rental-project

# 4    Results

We report results for the steps of the DSR process (Figure 1) in subchapters.

### 4.1 Problem identification, motivation, and objectives

The problem identification and motivation described in the introduction and background were translated into the following design objectives (DOs):
DO 1) Minimal risk for lender and renter, e.g., through non-returned NFTs
DO 2) Minimal cost for lender and renter, e.g., through opportunity costs due to (large) collaterals, high transactions fees for renter/lender contracts
DO 3) Value-based rental fees, e.g., based on positive perception, e.g., likes
DO 4) Easy access, e.g., ideally web-based and mobile friendly

### 4.2 Design and Development

**Design Decisions**: We made the following key design decisions. Blockchain technology was chosen since smart contracts allow to address the first three DOs. An assessment tool by Wüst and Gervais (2018), yielded the use of a public permissionless blockchain. Ethereum enjoys a large developer community and ecosystem (easier to find resources), security (It is time-proven and it has undergone multiple security audits), large support of token standards and the supply of popular NFTs is greatest[1]. There are alternatives such as Avalanche or Binance that offer higher throughput and lower transaction fees but come with disadvantages. Generally, they do not meet multiple of the above reasons. For example, Avalanche is less-established (smaller developer community) and newer with fewer security audits and less popular with fewer NFT supply. As the programming language for the backend, we used Solidity (Tikhomirov, 2018). We extended the commonly employed ERC-721 standard by OpenZeppelin (2022). To ensure easy access(DO4), we opted for a mobile-friendly web-based solution as frontend with TypeScript, which builds on top of JavaScript and allows to write type-safe and more readable code. To glue front- and backend, we used the well-established library "web3.js". The Angular Material library was used for UI components due to its ease of use and widespread adoption. We supported two types of wallet software to perform transactions on the blockchain, i.e., we chose MetaMask, which comes as a browser extension and WalletConnect, which is integrated into a mobile app. We opted for a collateral-free variant with a non-wrapped approach. This is because the collateral-free variants entail fewer risks for NFT owners and eliminate the barrier of high collateral costs for renters. Non-wrapped NFTs are considered more authentic, since they are associated directly with the original artwork (and not a copy like for wrapped NFTs). As value-based rental fees, we opted for pay-per-like, where the NFT owner defines the costs for a like. We deemed this more desirable than pay-per-view, since we explicitly aimed to capture positive reactions. To prevent click fraud, we decided to allow an increment only through an authorized party, which could be a service provider that aims to integrate the solution into their project. Commonly, this is the renter itself. The service provider must ensure that "likes" are genuine, e.g., non-fraudulent and non-duplicate. Thus, our solution does not aim to prevent click-fraud directly, but allows to defer it to another neutral entity that can employ any of the existing techniques for click-fraud. Both renters and lenders have an incentive to incorrectly detect non-genuine clicks (to pay less or charge more), while a separate entity per se has no such interest. To ensure the solvency of the renter to eliminate the risk that the renter is not paid due to excessive received likes, the renter must specify a limit with the maximum number of affordable likes and deposit the highest possible price in advance. The solution should then keep the amount until the rent deal is expired, and then the remaining money should be returned to the renter.

---

[1] https://opensea.io/rankings?sortBy=total_volume  (We checked for sales, e.g., in the last 30 days or all on 1st of June 2023)

**Development:** Next, we describe the creation of the artifactual solution as a Dapp including use cases that mostly directly translate to function calls (Figures 2 and 3).

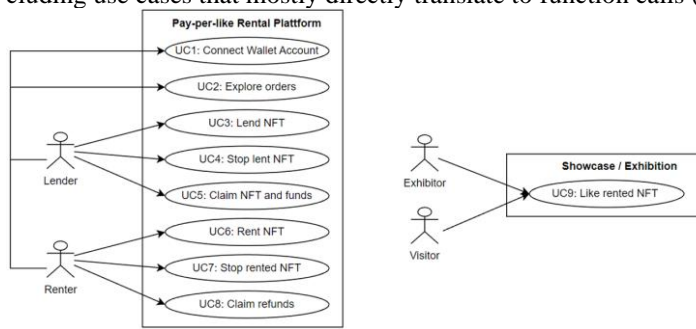

*Figure 2: Use Cases*

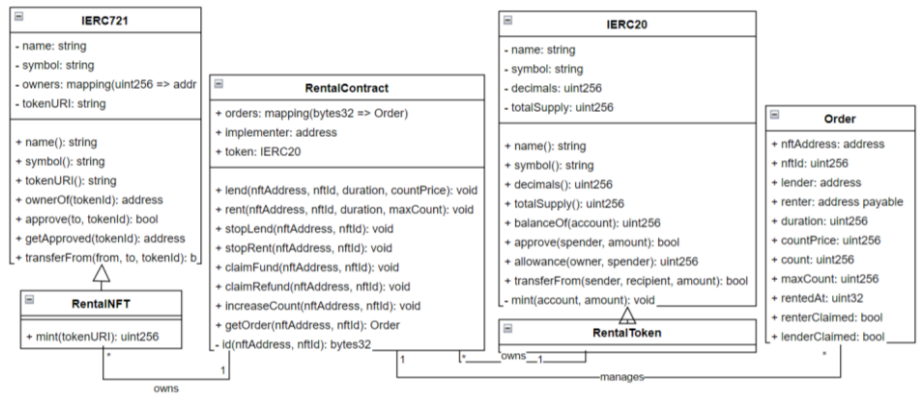

*Figure 3: Class Diagram*

Figure 2 depicts use cases with four human actors: A lender is a person aiming to rent NFTs. The lender opens a rental offer, i.e., an order, by specifying the maximum rent duration and the price per like. Once the lender has created an order, the NFT is transferred to the rental contract. If noone has rented the NFT, the lender can cease the order and reclaim the NFT. After an order is completed, the implementer reclaims the paid rental costs and the NFT. A renter is a person aiming to rent an NFT. The renter accepts a rental order by specifying a maximum rent duration and the maximum number of likes they are willing to pay for. To accept the order, they pay the maximum possible rent price, i.e., the maximum number of likes times the price per like. The order is either stopped by them or when the rent duration expires. After an order is completed, the final rent price is calculated, and the renter reclaims the remaining amount, which is calculated based on the difference between the already paid maximum rent price and the actual rent price.

The exhibitor is a service provider using the rental contract to enable rented NFTs within projects, e.g., to display on a webpage. The exhibitor verifies that an NFT is stored in the rental contract and the associated renter. The exhibitor also offers the functionality to "like" an item, which might trigger additional verification of users that are subject to the exhibitor. The "like" on a rented NFT is captured by incrementing the counter associated with the NFT on the rental contract. Only the exhibitor is authorized to increment this counter.

Use Case Descriptions: **Connect Wallet (UC 1)**: Lenders and renters want to use their accounts stored in their wallet, i.e., (exposed by) MetaMask or WalletConnect. The rental plattform provides a button to connect and disconnect their accounts. The connected account address is displayed in shortened form and associated with a unique profile image derived from the account address, also known as identicon.

**Explore Orders (UC 2)**: Lenders and renters want to have an overview of all lend and rent orders created by them. They can see information about the current RNT balance, the number of open lend orders, and the number of rented and lent NFTs.

**Lend NFT (UC 3)**: Lenders aim to lend their owned NFTs. The rental plattform allows to select an NFT asking for the maximum rent duration in days and the price per like in RNT. By clicking the lend button, the wallet appears and requests to accept or reject the NFT transfer from the lender's wallet to the rental (smart) contract. The rental contract stores the NFT until the order is completed and the lender reclaims it.

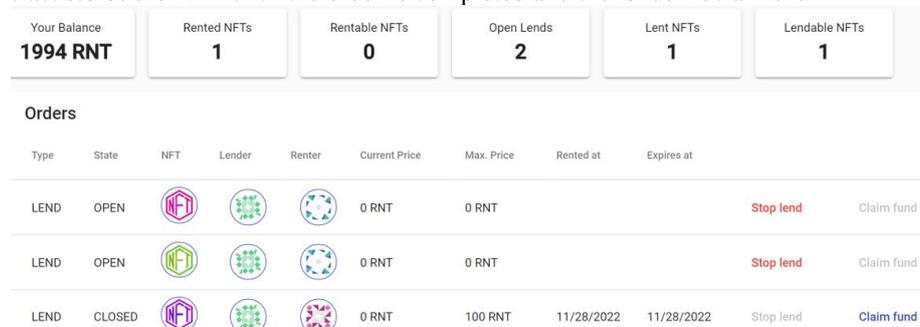

*Figure 4:Dashboard showing lend NFTs: Lending offers for NFTs that are not yet rented can be stopped, while funds can be claimed for completed rentals.*

**Stop Lent NFT(UC 4)**: The lender wants to stop an opened order which is not yet rented. User can click a "stop lend" which returns the NFT back to the lender's account. **Claim NFT and Funds(UC 5)**: The lender wants to reclaim the transferred NFT and the funds gained from the order. The lender has to accept or reject the gas fees required for the transaction. **Rent NFT(UC 6)**: The renter aims to rent a listed NFT. The renter selects an NFT and types the rent duration in days up to the maximum duration specified by the lender. The renter indicates the maximum like count they are willing to pay. The renter has to accept the RNT transfer from the lender's wallet to the rental contract.

**Stop Rented NFT(UC 7)**: The renter wants to cancel an already rented order. By clicking the "stop rent" button, the order is canceled, the renter can reclaim the refunds, and the lender can reclaim their NFT plus the funds gained from the rental deal.

**Claim Refunds(UC 8)**: The renter wants to claim the refunds. Upon clicking the "claim refunds" button, the user is asked to accept or reject the gas fees for the transaction.

**Like Rented NFT(UC 9)**: Visitors of an exhibition (or showcase) want to experience and like rented NFTs. The showcase provides access to rented NFTs and to like them, e.g., via a heart icon and a count. The exhibitor must validate clicks and allow the increment of the like counter.

## 4.3 Demonstration and Evaluation

For demonstration, the smart contracts implementing the use cases were deployed on a test network, called Goerli. We conducted system and unit tests. The end-to-end evaluation of the solution was executed via system tests. For each use case, the solution was operated manually and reviewed on Etherscan to assess whether the associated transactions had been successfully executed on the blockchain and yielded the correct balances for renter and lender and whether the system was easily usable. Figures 4 and 5 illustrate our solution.

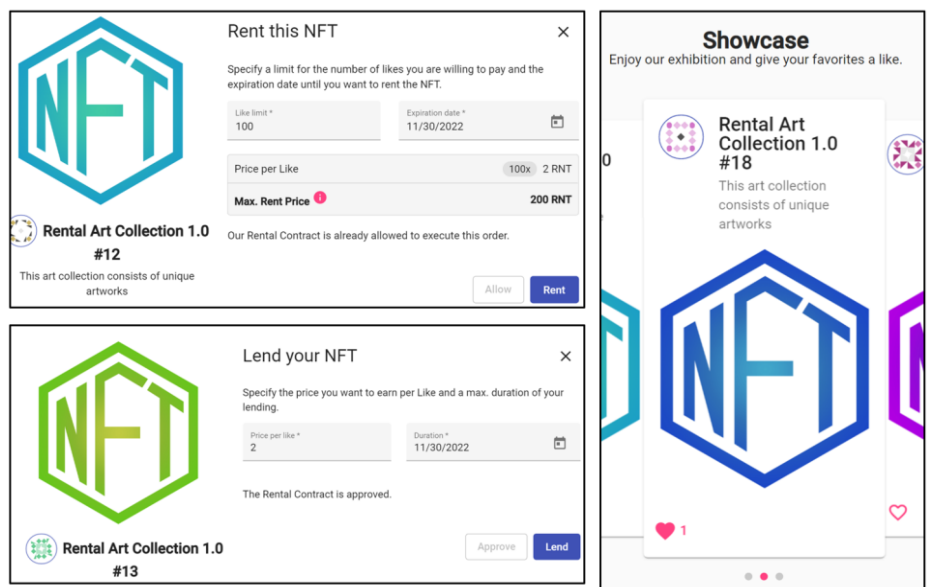

*Figure 5: Screenshots of rental (top left), lending(bottom left) and showcase with a heart button as "like" option*

Figure 6 shows unit tests for the renter. Overall, all 32 unit tests ran successfully. The tests are divided based on the perspectives of the deployer (which "installs" the solution), lender, renter and the exhibitor or implementer, i.e., which implements the actual exhibition and verifies like clicks. The unit tests achieved 100% coverage in all categories, namely statements, branches, functions, and lines, i.e., we also tested cases such as insufficient funds by a renter.

In order to understand the deployment and execution costs of the smart contracts, an analysis of the gas fees was performed on October 30, 2022 based on an ETH price of 1,597.70 USD and an average gas price of 7 gwei (=7 · $10^{-9}$ ETH). The Hardhat library retrieves the current Ethereum price and the estimated gas price from the mainnet by fetching the data from the crypto asset price tracking website "CoinMarketCap". Hardhat then analyzes how much gas in gwei the function requires to be executed on the Ethereum blockchain and calculates costs. Execution costs for functions listed in Figure 3 varied between 0.45 and 2.06 USD. The deployment cost for our smart contract was $28.08, which a service provider must initially pay when integrating the rental solution.

We conclude the evaluation with respect to our design objectives:

DO1(Minimal risk for lender and renter): Using smart contracts, it is automatically enforced that all NFTs are returned and a lender is paid. The renter in turn can prove authenticity of NFTs due to transparency of the blockchain. We defer the detection and handling of click fraud to the exhibitor. An exhibitor can control whom to pay. This is fine as long as the exhibitor is an external service provider who favors neither the renter nor the lender. Otherwise, there is a risk of click fraud, either by not recording clicks (in case the renter is favored) or by not properly detecting fraud by others (to favor the lender).

DO2(Minimal cost for lender and renter): Our solution avoids the use of (large) collaterals, which minimizes interest and opportunity costs due to inaccessible assets. Furthermore, our solution is fully digital and, therefore, should be highly cost-efficient. Still, fees are significant. Total costs due to gas fees for deployment are about $28. Executing a lend or rental order on the smart contact costs between $2 and $3. While this is a considerable amount, it is still little compared taking into account that no intermediary is needed and no collateral. However, "a like", i.e., increasing the counter (executing function increaseCount in Figure 3 cost about $0.45. This is a considerable amount and well beyond what is typically paid in online-advertising settings. We elaborate on mitigations (and their trade-offs) in the discussion.

DO3(Value-based rental fees): Our pay-per-like scheme allows a value-based assessment. However, it is biased towards positive feedback, i.e., there is no "dislike" button. Furthermore, preventing click fraud is not solved out-of-the-box using a blockchain. We present further options in the discussion.

DO4(Easy access): Our solution is easy to access using a web-browser and mobile devices. But interactions with Dapps require additional tools, such as wallet software, which is not always supported, e.g., in older browsers.

```
Renter
  ✓ should execute rent transaction (112ms)
  ✓ should not execute rent transaction with a duration <= 0 (49ms)
  ✓ should not execute rent transaction with a max count e <= 0
  ✓ should not execute rent transaction when renter has not enough balance (144ms)
  ✓ should not execute rent transaction when nft is already rented
  ✓ should not execute increaseCount
  ✓ should execute stop rent transaction (45ms)
```

*Figure 6: Subset of unit tests. Overall, all 32 tests covering 100% of code ran successfully*

## 5 Discussion and Future Work

In our evaluation, we already pointed out some strengths and limitations of the proposed solution. Overall, blockchain comes with many advantages and our design objectives could mostly be addressed. One challenge of blockchain is that **technology** is still **rapidly evolving**. For example, the (new) ERC-4907 (rentable tokens) offers to grant a time-limited role with restricted rights to a token. As of the writing of this paper, it is not used, since there is no official and reviewed implementation of the interfaces. While it comes with benefits, it remains uncertain whether this standard (potentially well-suited for renting solutions) will find adaptations in the NFT community, making technology decisions difficult.

Another issue not resolved fully with blockchain is dealing with click-fraud. Our solution relies on the exhibitor to ensure that click fraud is detected and handled. That is, it enables the **use of third parties to address fraud**. This is beneficial since the renter and lender in principle do not trust each other, since both have opposite incentives when it comes to counting likes. However, both the renter and lender do not obtain any indicators that might be used to assess if fraud occurred, i.e., they must still trust the third party. In principle, any click on the "like" button (irrespective of whether it is deemed fraudulent or not) could be stored on the blockchain with all kinds of meta-data, such as if the click was deemed fraudulent or not. This might allow verifying the correctness of like classifications as fraudulent or not. However, this would increase gas fees dramatically, even, using just a simple counter that records any "like" clicks could be problematic, since a bot could trigger costs in the form of gas fees with every click. Thus, fraud remains an issue. We observe **a trust-cost tradeoff**. We define the trust costs as the additional costs that occur to capture information that allows to verify actions of another party. In our case, costs are gas fees that arise to capture information on fraud decisions by the exhibitor. The third party could use the same techniques for fraud detection that are currently employed already in the context of online ad click fraud, e.g., (Zhang and Guan, 2008). Such detection techniques can rely on behavioral information or require user authentication. On many online platforms, such as Youtube and StackOverflow, only users with a verified account can like or add comments.

The pay-per-like pricing might not always be preferred, i.e., one might prefer views or other metrics. Our **implementation is generic**; the counters implemented on the smart contract could also be used to count different actions like views, points, or anything else. However, care must be taken with respect to costs. For example, recording views is likely much more expensive than likes, since views are usually at least an order of magnitude larger than likes, i.e., for a large number of views gas fees would also be an order of magnitude larger for counting likes. Gas fees are already problematic when counting likes, i.e., they are about 0.5 USD. **Gas fees can be prohibitive** in many settings. One way to address this issue is to **increment the like counter from time to time**, e.g., only after every 10 verified like clicks. However, even using such ideas, our solution occurs significant gas fees, although all smart contracts are rather simple. Another option to reduce fees is to use a chain with cheaper transaction costs such as Avalanche or Binance instead of Ethereum. However, they tend to have drawbacks com-pared to Ethereum such as higher costs for implementation (due to lower online resources, fewer standards), less popularity (fewer NFTs), lower security (fewer validators). Furthermore, even if costs are about a factor 20 or so lower than on Ethereum as is roughly the case for Binance[2] , they are still a concern - especially if value-based pricing models such as pay-per-view are used. As such, using a blockchain-based approach is not preferable for "cheap" artwork, i.e., taking risks might be more economical than paying fees. Even worse gas fees can exceed potential income. This is problematic, since niche or upcoming artists renting their work do not benefit from the technology in the same way as well-established artists with expensive works, since gas fees take relatively a much larger share of their earnings. That is, for expensive rentals blockchain constitutes a major facilitator, since risks are mitigated and access is simplified. While for artwork that cannot be lend for high prices, blockchains like Ethereum are not an option as gas fees exceed rental income. Thus, **blockchain solutions appear unfair and foster injustices**, i.e., making the rich richer and the poor poorer and contributing to a reduction of cultural diversity. Similar criticism is well-established for other technologies. For example, artificial intelligence is known for its dark sides – especially biases (Meske *et al.*, 2022) and ability to deceive (Schneider, Vlachos and Meske, 2022). For blockchain, to this end only the unfairness of mining mechanisms and rules for proof-of-work based blockchains such as Bitcoin have been mentioned (Ferdous, Chowdhury and Hoque, 2021). Generally, it can be said that small transactions are not profitable so that certain pricing models (like pay-per-view) or market participants who cannot charge high rental fees (like niche artists) are excluded. This is in contrast to ordinary centralized web platforms, i.e., Web2, where computations or transactions occur little cost. We call this the **dark sides of large blockchain fees**. Centralized solutions do not occur any coordination costs in contrast to decentralized approaches. Thus, Dapps are inherently inferior from a computational perspective as witnessed by many papers in the field of distributed computing analyzing the costs of decentralized algorithms, e.g., (Harris, Schneider and Su, 2018), especially, consensus algorithms(Chaudhry and Yousaf, 2018).

As future work to improve the current solution one can include chains with lower costs and fraud detection, e.g., by authentication via a blockchain (Patel *et al.*, 2019).

## 6    Conclusions

We have presented a value-based rental solution for NFTs. Our study showed that blockchain technology has many inherent properties that make it a preferable technology for such a problem, e.g., smart contracts that ensure payments or the return of items without the need for collatorals. However, for handling fraud trust still relies on a third party and we observe a trust-cost tradeoff, i.e., larger costs for more trust. From a societal level, blockchain technology with high fees can appear unfair as it benefits wealthy artists more than non-wealthy ones.

---

[2] https://academy.binance.com/en/articles/binance-smart-chain-vs-ethereum-what-s-the-difference   We use Binance's on report as a lower bound for Binance fees. They state average transactions costs of 2.46$ for Binance and 68.72$ for Etheruem. Prices fluctuate strongly.

# References

Antonopoulos, A.M. and Wood, G. (2019) *Mastering Ethereum: Building smart contracts and DApps*. First edition. O'Reilly.

Bauer, J.M. and Wildman, S.S. (2012) 'The economics of usage-based pricing in local broadband markets'. Available at: https://techliberation.com/wp-content/uploads/2012/12/wildmanreport_web.pdf.

Bennion, Jackie (2021) 'VP Bank "Pioneering" Tokenization Trend'. Available at: https://www.wealthbriefing.com/html/article.php?id=193079 (Accessed: 10 February 2022).

BoredJobs (2022) 'Hire the perfect NFT for your brand'. Available at: https://www.boredjobs.com/about.

vom Brocke, J. *et al.* (2018) 'Own–The Case of a Blockchain Business Model Disrupting the Equity Market', *Controlling30*, (5), pp. 19–25.

Buterin, V. (2014) 'Ethereum: A Next-Generation Smart Contract and Decentralized Application Platform.' Available at: https://ethereum.org/669c9e2e2027310b6b3cdce6e1c52962/Ethereum_Whitepaper_-_Buterin_2014.pdf.

Chaudhry, N. and Yousaf, M.M. (2018) 'Consensus algorithms in blockchain: Comparative analysis, challenges and opportunities', in *2018 12th International Conference on Open Source Systems and Technologies (ICOSST)*. IEEE, pp. 54–63.

Ding, C. *et al.* (2017) 'The power of the "like" button: The impact of social media on box office', *Decision Support Systems*, 94, pp. 77–84.

Dr. Wood, G. (2022) 'Ethereum Yellow Paper: a formal specification of Ethereum, a programmable blockchain'. Available at: https://ethereum.github.io/yellowpaper/paper.pdf.

Ethereum Community (2022a) 'ERC-721 Non-Fungible Token Standard'. Available at: https://ethereum.org/en/developers/docs/standards/tokens/erc-721/.

Ethereum Community (2022b) 'Ethereum Accounts'. Available at: https://ethereum.org/en/developers/docs/accounts/.

Ferdous, M.S., Chowdhury, M.J.M. and Hoque, M.A. (2021) 'A survey of consensus algorithms in public blockchain systems for crypto-currencies', *Journal of Network and Computer Applications*, 182, p. 103035.

Harris, D.G., Schneider, J. and Su, H.-H. (2018) 'Distributed ($\Delta+ 1$)-Coloring in Sublogarithmic Rounds', *Journal of the ACM (JACM)*, 65(4), pp. 1–21.

Hevner *et al.* (2004) 'Design Science in Information Systems Research', *MIS Quarterly*, 28(1), p. 75. Available at: https://doi.org/10.2307/25148625.

Hirai, Y. (2017) 'Defining the Ethereum Virtual Machine for Interactive Theorem Provers', in M. Brenner et al. (eds) *Financial Cryptography and Data Security*, pp. 520–535. Available at: https://doi.org/10.1007/978-3-319-70278-0_33.

IQSpace (2022) 'What is the IQ Space?' Available at: https://docs.iq.space/nft/general/iq-space.

Ives, N. (2005) 'Web Marketers Fearful of Fraud in Pay-Per-Click'. The New Yourk Times. Available at: https://www.nytimes.com/2005/03/03/business/media/web-marketers-fearful-of-fraud-in-payperclick.html.

Khraim, H.S. and Alkrableih, A.A. (2015) 'The Effect of Using Pay Per Click Advertisement on Online Advertisement Effectiveness and Attracting Customers in E-marketing Companies in Jordan', *International Journal of Marketing Studies*, 7(1). Available at: https://doi.org/10.5539/ijms.v7n1p180.

Kirli, D. *et al.* (2022) 'Smart contracts in energy systems: A systematic review of fundamental approaches and implementations', *Renewable and Sustainable Energy Reviews*, 158, p. 112013.

Meske, C. *et al.* (2022) 'Explainable artificial intelligence: objectives, stakeholders, and future research opportunities', *Information Systems Management*, 39(1), pp. 53–63.

OpenZeppelin (2022) 'OpenZeppelin Docs: A library for secure smart contract development.' Available at: https://docs.openzeppelin.com/contracts/4.x/.

Patel, S. *et al.* (2019) 'DAuth: A decentralized web authentication system using Ethereum based blockchain', in *2019 International Conference on Vision Towards Emerging Trends in Communication and Networking (ViTECoN)*. IEEE, pp. 1–5.

Peffers, K. *et al.* (2007) 'A design science research methodology for information systems research', *Journal of management information systems*, 24(3), pp. 45–77.

Regner, F., Urbach, N. and Schweizer, A. (2019) 'NFTs in practice–non-fungible tokens as core component of a blockchain-based event ticketing application', in *International Conference on Information Systems (ICIS)*.

reNFT (2022) 'Rental Infrastructure for the Metaverse'. Available at: https://medium.com/renftlabs/renft-announces-new-funding-round-to-accelerate-growth-of-gamefi-in-web3-830702329085.

Rosso, M.A. and Jansen, B.J. (2010) 'Brand Names as Keywords in Sponsored Search Advertising', in *Communications of the Association for Information Systems*.

Schneider, J., Vlachos, M. and Meske, C. (2022) 'Deceptive AI explanations: Creation and detection', in *International Conference on Agents and Artificial Intelligence (ICAART)*.

Solorio, K., Kanna, R. and Hoover, D.H. (2019) *Hands-on smart contract development with Solidity and Ethereum: From fundamentals to deployment*. First edition. O'Reilly Media Inc.

Szabo, N. (1994) 'Smart Contracts'. Available at: https://www.fon.hum.uva.nl/rob/Courses/InformationInSpeech/CDROM/Literature/LOTwinterschool2006/szabo.best.vwh.net/smart.contracts.html.

Tikhomirov, S. (2018) 'Ethereum: state of knowledge and research perspectives', in *International Symposium of Foundations and Practice of Security*, pp. 206–221.

Vom Brocke, J. *et al.* (2020) 'Accumulation and evolution of design knowledge in design science research: a journey through time and space', *Journal of the Association for Information Systems*, 21(3), p. 9.

Voshgmir, S. (2018) 'Fungible Tokens vs. Non-Fungible Tokens.' Available at: https://blockchainhub.net/blog/blog/nfts-fungible-tokens-vs-non-fungible-tokens/.

Weinhardt, C. *et al.* (2009) 'Business Models in the Service World', *IT Professional*, 11(2), pp. 28–33. Available at: https://doi.org/10.1109/MITP.2009.21.

Werner, F. *et al.* (2021) 'Blockchain adoption from an interorganizational systems perspective–a mixed-methods approach', *Information Systems Management*, 38(2), pp. 135–150.

Wüst, K. and Gervais, A. (2018) 'Do you need a blockchain?', in *Crypto valley conference on blockchain technology (CVCBT)*, pp. 45–54.

Zhang, L. and Guan, Y. (2008) 'Detecting click fraud in pay-per-click streams of online advertising networks', in *International Conference on Distributed Computing Systems*, pp. 77–84.

Zheng, G. *et al.* (2021) *Ethereum Smart Contract Development in Solidity*. 1st edition. Available at: https://doi.org/10.1007/978-981-15-6218-1.